\documentclass[twocolumn,showpacs,superscriptaddress,prb,aps,floatfix]{revtex4-2}
\usepackage{graphicx}
\usepackage{rotating}
\usepackage{amsmath}
\usepackage[usenames,dvipsnames]{color}
\usepackage{float}
\usepackage{rotating}
\usepackage{booktabs}
\usepackage{multirow}
\usepackage{longtable}
\usepackage[colorlinks,linkcolor=blue,citecolor=blue]{hyperref}
\usepackage[T1]{fontenc}
\usepackage{bm}
\usepackage{upgreek}
\usepackage{comment}
\usepackage{amsmath,relsize,graphicx}
\usepackage{booktabs}
\makeatletter

\newcommand{\Rmnum}[1]{\expandafter\@slowromancap\romannumeral #1@}

\newcommand{\DC}[1]{\textcolor{red}{To Domenico and Cesare:}}
\makeatletter
\hfuzz=\maxdimen
\begin{document}

\title{Thermoelectric Properties of Copper-based Chalcopyrite Semiconductors Cu$MX_2$ ($M$ = Al, Ga, and In; $X$ = S, Se, and Te) from First-Principles Calculations}

\author{Wu Xiong}
\affiliation{Key Laboratory of Advanced Materials and Devices for Post-Moore Chips, Ministry of Education, University of Science and Technology Beijing, Beijing 100083, China}
\affiliation{School of Mathematics and Physics, University of Science and Technology Beijing, Beijing 100083, China}

\author{Zhonghao Xia}
\affiliation{Key Laboratory of Advanced Materials and Devices for Post-Moore Chips, Ministry of Education, University of Science and Technology Beijing, Beijing 100083, China}
\affiliation{School of Mathematics and Physics, University of Science and Technology Beijing, Beijing 100083, China}

\author{Zhongjuan Han}
\affiliation{Key Laboratory of Advanced Materials and Devices for Post-Moore Chips, Ministry of Education, University of Science and Technology Beijing, Beijing 100083, China}
\affiliation{School of Mathematics and Physics, University of Science and Technology Beijing, Beijing 100083, China}

\author{Dong Yao}
\affiliation{School of Mathematics and Physics, University of Science and Technology Beijing, Beijing 100083, China}

\author{Jiangang He}
\email{jghe2021@ustb.edu.cn}
\affiliation{Key Laboratory of Advanced Materials and Devices for Post-Moore Chips, Ministry of Education, University of Science and Technology Beijing, Beijing 100083, China}
\affiliation{School of Mathematics and Physics, University of Science and Technology Beijing, Beijing 100083, China}

\date{\today}

	\begin{abstract}
	 Copper-based chalcopyrite semiconductors have attracted sustained interest owing to their promising thermoelectric (TE) performance, yet the microscopic origins of their TE behavior remain incompletely understood. Here, we systematically investigate the TE properties of Cu$MX_2$ ($M=$ Al, Ga, and In; $X=$ S, Se, and Te) using first-principles calculations. Electronic transport is computed from explicit electron--phonon coupling based on Perdew--Burke--Ernzerhof (PBE) functional calculated band structures including spin--orbit coupling (SOC), with band gaps and dielectric constants corrected using the screened hybrid functional HSE06. Lattice thermal conductivities ($\kappa_{\mathrm{L}}$) are obtained by solving the phonon Boltzmann transport equation with three- and four-phonon scattering, using temperature-renormalized second-order force constants from 300 to 800~K. For $p$-type doping, the calculated electrical conductivities ($\sigma$), hole mobilities ($\mu$), Seebeck coefficients ($S$), and power factors (PFs) of CuGaTe$_2$ and CuInTe$_2$ show excellent agreement with experimental data. At fixed temperature and hole concentration, as $X$ varies from S to Te, the hole mobility increases markedly due to progressively weaker polar--optical--phonon scattering, reflecting the reduced ionic contribution to the dielectric response in compounds with heavier chalcogens. Combined with smaller transport effective masses, Cu$M$Te$_2$ compounds therefore exhibit high $\sigma$ and large PFs. Across the Cu$MX_2$ family, the anomalously lower $\kappa_{\mathrm{L}}$ of Cu$M$Se$_2$ relative to Cu$M$Te$_2$ arises primarily from enhanced three-phonon scattering at low-frequency region. For a given $M$, Cu$M$S$_2$ displays the steepest temperature-induced decrease in $\kappa_{\mathrm{L}}$ and attains a smaller $\kappa_{\mathrm{L}}$ than Cu$M$Se$_2$ and Cu$M$Te$_2$ at 800~K. Given the low band degeneracy and comparatively modest hole mobilities of Cu$MX_2$ compounds, the most effective routes to further improve their TE performance are to enhance $\sigma$ and reduce $\kappa_{\mathrm{L}}$ through doping.
	 
	\end{abstract}

	\maketitle
	\section{INTRODUCTION}
    Thermoelectric energy conversion is one of the most attractive sustainable energy technologies because it enables the direct, reversible conversion between heat and electricity in all-solid-state devices, supporting both power generation and cooling. This capability offers substantial opportunities for waste-heat recovery and for on-chip or cryogenic refrigeration, and has consequently sustained significant interest in the materials community~\cite{science.1159725,2008Complex,2022Chemical}. The performance of a TE material is quantified by the dimensionless figure of merit $ZT=\sigma S^{2}T/\kappa$, where $S^{2}\sigma$ is PF and the total thermal conductivity $\kappa=\kappa_{\mathrm{L}}+\kappa_{\mathrm{e}}$ contains lattice ($\kappa_{\mathrm{L}}$) and electronic ($\kappa_{\mathrm{e}}$) contributions. A central challenge in enhancing $ZT$ is the intrinsic interdependence of $\sigma$ and $S$ on the carrier concentration and the band structure, which constrains improvements in PF. For example, insulators and lightly doped semiconductors typically exhibit large $S$ but very low $\sigma$ due to the paucity of carriers~\cite{2008Complex}, whereas heavily doped semiconductors or metals often show high $\sigma$ but small $S$. In general, large $S$ benefits from a large density-of-states effective mass, while high $\sigma$ prefers a small transport effective mass and high mobility. Consequently, band-structure engineering strategies such as band convergence and increased valley degeneracy are favored, as they help maintain a large $S$ while enabling high $\sigma$~\cite{2008Complex,https://doi.org/10.1002/adma.201202919,https://doi.org/10.1002/anie.201508381,he2019designing,xiong2025forbidden}. Because $\kappa_{\mathrm{e}} \approx \mathrm{L} \,\sigma T$ by the Wiedemann--Franz law (with Lorenz number $\mathrm{L}$)~\cite{solidstatephysics}, reducing $\kappa_{\mathrm{L}}$ is often the most effective and relatively independent route to increasing $ZT$. Within simple kinetic theory~\cite{tritt2005thermal}, $\kappa_{\mathrm{L}}=\tfrac{1}{3} C_{\mathrm{V}} v_{\mathrm{g}}^{2}\tau$, where $C_{\mathrm{V}}$ is the heat capacity, $v_{\mathrm{g}}$ the phonon group velocity, and $\tau$ the phonon relaxation time. The group velocity scales roughly with bond stiffness and inversely with the average atomic mass. Accordingly, materials with short $\tau$, soft bonding~\cite{https://doi.org/10.1002/adfm.202108532,https://doi.org/10.1002/advs.202417292}, and heavy constituent atoms tend to exhibit low $\kappa_{\mathrm{L}}$. Effective strategies to reduce $\tau$ include point-defect and alloy-disorder engineering~\cite{Mao03042018}, nanostructured precipitates~\cite{doi:10.1126/science.1092963,2012High,doi:10.1126/science.1156446}, lone-pair-induced anharmonicity~\cite{PhysRevLett.107.235901,2013Lone}, and rattling modes~\cite{koza2008breakdown,2015Impact,2016Ultralow}.

    Over the past few decades, substantial progress in TE materials has spurred the discovery and optimization of numerous high-$ZT$ systems, including half-Heusler alloys~\cite{chen2013recent,fu2015realizing}, IV--VI semiconductors~\cite{zhao2014ultralow,jiang2022high}, clathrates~\cite{B916400F,nolas1998semiconducting}, skutterudites~\cite{nolas1999skutterudites,koza2008breakdown,tang2015convergence}, Zintl phases~\cite{zheng2020ternary,https://doi.org/10.1002/adfm.201000970}, and copper-based chalcogenides~\cite{https://doi.org/10.1002/idm2.12134,https://doi.org/10.1002/adfm.202108532}. Among these, chalcopyrites with composition $AMX_2$, which are structural derivatives of the zinc-blende lattice, exhibit outstanding TE performance owing to their high PFs and comparatively low $\kappa_{\mathrm{L}}$~\cite{https://doi.org/10.1002/adma.201400058,https://doi.org/10.1002/idm2.12134,CuGaTe2,CuInTe2,10.1063/1.3678044,2020Origin} For example, CuGaTe$_2$ achieves a peak $ZT$ of 1.4 at 950~K~\cite{CuGaTe2}, while CuInTe$_2$ reaches 1.18 at 950~K~\cite{CuInTe2}. These benchmarks highlight the strong potential of Cu-based ternary chalcopyrites for TE applications. Despite this promise, research has predominantly focused on CuGaTe$_2$ and CuInTe$_2$~\cite{CuInTe2,2016Simultaneous,D1TA02893F,WANG2024151588,2020Origin,CuGaTe2,2017Substitutional,C6TA06033A,CuAg2024}, leaving other compositions relatively underexplored~\cite{CuAlTe,CuAlS,CuInS}. Moreover, existing theoretical studies often lack accuracy and systematic cross-material comparisons: SOC is frequently omitted in electronic-structure and transport calculations, and charge transport is commonly treated using the deformation-potential model~\cite{PhysRev.80.72,PhysRev.101.944} or other simplified approximations to electron--phonon interactions~\cite{ganose2021efficient}. In addition, many investigations are limited to phonon transport properties~\cite{CuAlTe,CuAg2024,XUE2014143,CHEN2019369,theory2,plata2022charting,PhysRevB.106.094317,10.1063/5.0216813,PhysRevB.105.245204}, without a comprehensive, first-principles treatment of both electronic and lattice contributions to $ZT$.

    In this work, we systematically investigate the TE properties of Cu-based chalcopyrite compounds, Cu$MX_2$ ($M$ = Al, Ga, and In; $X$ = S, Se, and Te), using first-principles calculations. The electronic transport properties are evaluated by explicitly including electron-phonon coupling, based on electronic structures obtained with the PBE functional and incorporating spin--orbit coupling (SOC), with band gaps and dielectric constants corrected using the HSE06 functional. The $\kappa_{\mathrm{L}}$ are determined by solving the phonon Boltzmann transport equation, taking into account both three- and four-phonon scattering processes, with renormalized second-order force constants computed at all studied temperatures. Our results demonstrate that SOC significantly impacts the band degeneracy at the valence band maximum for CuAlTe$_2$, CuGaSe$_2$, CuGaTe$_2$, CuInSe$_2$, and CuInTe$_2$, thereby influencing their electronic transport properties. At a given temperature and hole concentration, the hole mobility of Cu$MX_2$ compounds increases substantially as $X$ varies from S to Te, resulting in a corresponding enhancement of $\sigma$. This trend is attributed to the progressively weaker polar optical phonon scattering, which arises from the decreasing ionic contribution to the dielectric constants as the atomic mass of $X$ increases. Combined with their smaller transport effective masses, Cu$M$Te$_2$ compounds thus exhibit large PFs. Our calculated values for hole mobility, $\sigma$, and $S$ are in good agreement with experimental data for CuGaTe$_2$ and CuInTe$_2$~\cite{CuGaTe2,CuInTe2,2020Origin}. Phonon and $\kappa_{\mathrm{L}}$ calculations indicate that three-phonon scattering plays a more significant role than four-phonon processes in suppressing heat transport in these materials. The anomalously low $\kappa_{\mathrm{L}}$ observed in Cu$M$Se$_2$ compounds is explained by their high three-phonon scattering rates. Our computed $ZT$ value for CuInTe$_2$ (0.71) is in excellent agreement with the experimental value (0.74)~\cite{2020Origin}. The highest $ZT$ values for CuAlS$_2$ and CuGaS$_2$ are 0.76 and 0.74, respectively, with the maximum $ZT$ among all studied compounds achieved by CuAlS$_2$ at 800~K. Given the low band degeneracy and relatively modest hole mobility in this family, additional reductions in $\kappa_{\mathrm{L}}$ and enhancements in $\sigma$ through doping represent promising strategies for further improving the TE performance of Cu$MX_2$ compounds.

	\begin{figure}[th!]
		\centering
		\includegraphics[width=1.0\linewidth]{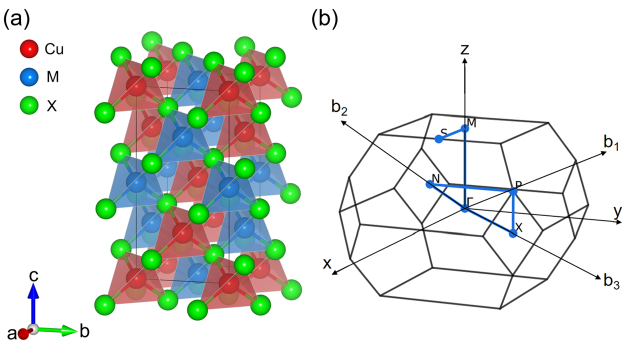}
		\caption{(a) and (b) depict the crystal structure and the first Brillouin zone of Cu$MX_2$ ($M$ = Al, Ga, and In; $X$ = S, Se, and Te), respectively. In these illustrations, red, blue, and green spheres represent Cu, $M$, and $X$ atoms, respectively.}
		\label{crystal structure}
	\end{figure}

	\begin{table*}
		\scriptsize
		\setlength{\tabcolsep}{3.8pt} 
		\renewcommand{\arraystretch}{1.3} 
		\centering
		\caption{Comparison of our calculated lattice constants ($a$, $c$), tetragonal distortion parameter ($\eta$), and $X$-Cu-$X$ bond angle for Cu$MX_2$ ($M=$ Al, Ga, and In; $X=$ S, Se, and Te), computed with and without spin–orbit coupling (W SOC and W/O SOC), against experimental values (Exp.) reported in Ref.~\onlinecite{1953}.}
		\begin{tabular}{ccccccccccccccc}
			\hline
			\hline
  &\multicolumn{3}{c}{$a$ (\AA)}&\multicolumn{3}{c}{$c$ (\AA)}&\multicolumn{3}{c}{$\eta$}&\multicolumn{3}{c}{$\angle(X\mathrm{-Cu-}X)$ ($^{\circ}$)} \\               
            \cmidrule(l{6pt}r{6pt}){2-4}     \cmidrule(l{6pt}r{6pt}){5-7}   \cmidrule(l{6pt}r{6pt}){8-10} \cmidrule(l{6pt}r{6pt}){11-13}
    Comp.  & W SOC & W/O SOC&  Exp.  & W SOC  & W/O SOC& Exp.  & W SOC & W/O SOC&  Exp. & W SOC  & W/O SOC &     Exp.      \\
\hline
CuAlS$_2$  & 5.291 & 5.291  & 5.31   & 10.471 & 10.471 & 10.42 & 0.990 & 0.989  & 0.981 & 108.87 & 110.68  & 108.87, 110.68 \\
CuAlSe$_2$ & 5.598 & 5.598  & 5.60   & 11.053 & 11.052 & 10.90 & 0.987 & 0.987  & 0.973 & 109.07 & 110.27  & 109.07, 110.28 \\
CuAlTe$_2$ & 6.043 & 6.039  & 5.96   & 12.007 & 11.999 & 11.78 & 0.993 & 0.993  & 0.988 & 108.88 & 109.77  & 108.91, 109.75 \\
CuGaS$_2$  & 5.317 & 5.317  & 5.34   & 10.529 & 10.528 & 10.47 & 0.990 & 0.990  & 0.980 & 109.34 & 109.73  & 109.34, 109.74 \\
CuGaSe$_2$ & 5.610 & 5.609  & 5.60   & 11.128 & 11.126 & 10.91 & 0.992 & 0.992  & 0.974 & 109.24 & 109.60  & 109.21, 109.60 \\
CuGaTe$_2$ & 6.033 & 6.028  & 5.99   & 12.045 & 12.032 & 11.91 & 0.998 & 0.998  & 0.994 & 108.15 & 110.14  & 108.20, 110.11 \\
CuInS$_2$  & 5.529 & 5.530  & 5.51   & 11.147 & 11.148 & 11.06 & 1.008 & 1.008  & 1.004 & 105.71 & 111.39  & 105.70, 111.39 \\
CuInSe$_2$ & 5.814 & 5.816  & 5.77   & 11.710 & 11.700 & 11.55 & 1.007 & 1.006  & 1.001 & 105.56 & 111.46  & 105.63, 111.42 \\
CuInTe$_2$ & 6.243 & 6.238  & 6.16   & 12.534 & 12.518 & 12.34 & 1.004 & 1.003  & 1.002 & 105.38 & 111.55  & 105.46, 111.52 \\
\hline
\end{tabular}
\label{lattice constant}
\end{table*}  

    \begin{figure*}[th!]
	\centering
	\includegraphics[width=1.0\linewidth]{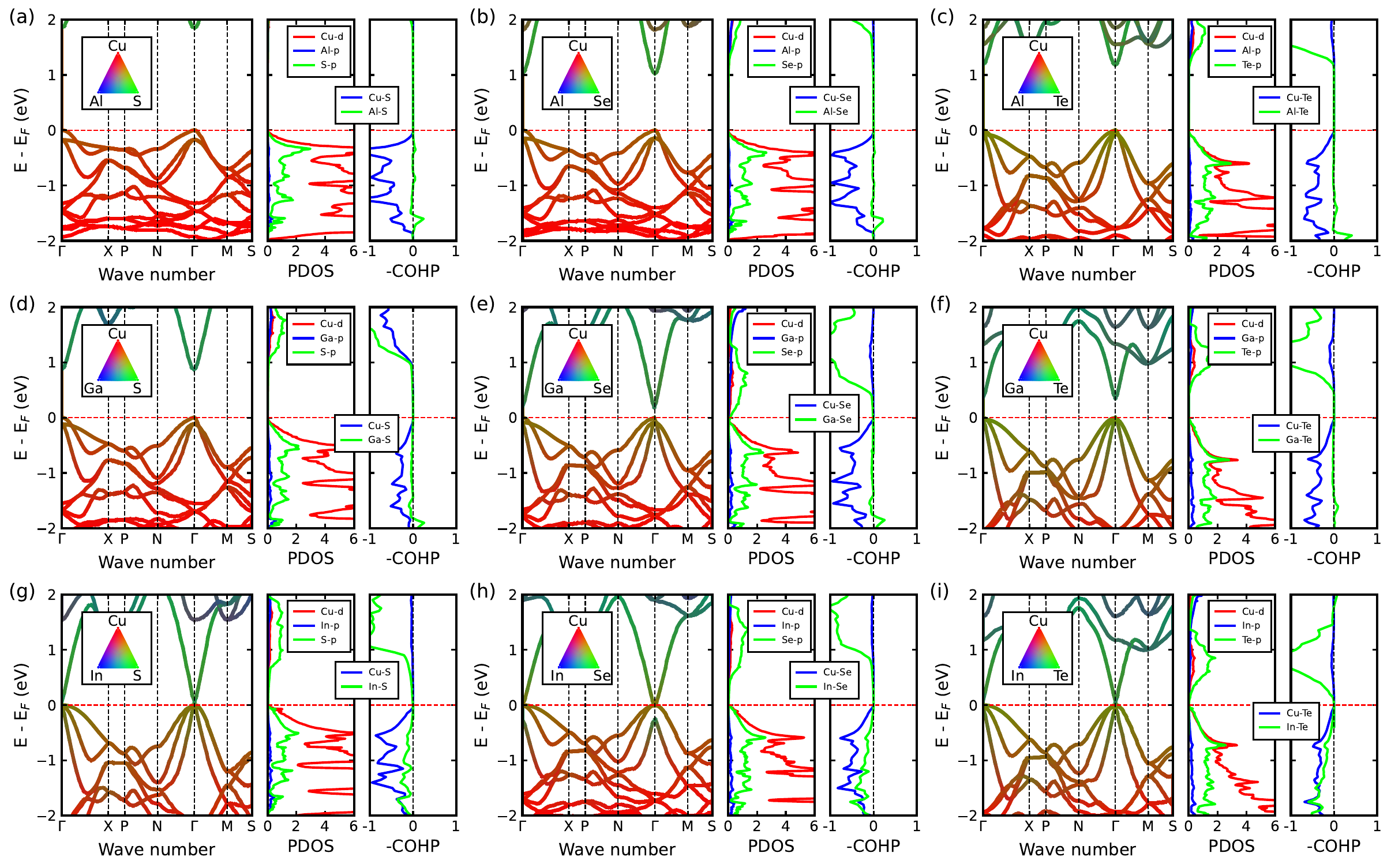}
	\caption{Atom-projected band structures, PDOS, and -COHP analyses for Cu$MX_2$ ($M=$ Al, Ga, and In; $X=$ S, Se, and Te). (a)-(i) correspond to CuAlS$_2$, CuAlSe$_2$, CuAlTe$_2$, CuGaS$_2$, CuGaSe$_2$, CuGaTe$_2$, CuInS$_2$, CuInSe$_2$, and CuInTe$_2$, respectively.}
	\label{bandnosoc}
    \end{figure*}

	\maketitle
	\section{COMPUTATIONAL METHODS}
	Electronic-structure calculations were performed within density functional theory (DFT)~\cite{dft1,dft2} as implemented in the Quantum ESPRESSO (QE) package~\cite{QE}. Fully relativistic, optimized norm-conserving Vanderbilt (ONCV) pseudopotentials~\cite{ONCV} were employed, and exchange--correlation effects were treated using the Perdew--Burke--Ernzerhof (PBE) functional of generalized-gradient approximation (GGA)~\cite{PBE,GGA}. A plane-wave basis with kinetic-energy cutoff of 70~Ry was used. Total energies and forces were converged to $10^{-6}$~Ry and $10^{-6}$~Ry/Bohr, respectively. Structural relaxations and self-consistent field (SCF) calculations employed a $6\times6\times6$ $k$-point mesh. Convergence with respect to the cutoff energy and $k$-point sampling was verified (see Fig.~S2). Because semilocal functional PBE normally underestimates band gaps and tends to overestimate the high-frequency dielectric constant ($\varepsilon^\infty$), the band gaps and $\varepsilon^\infty$ were evaluated using the Heyd--Scuseria--Ernzerhof screened hybrid functional (HSE06)~\cite{hse.10.1063/1.1564060} as implemented in the Vienna \textit{ab initio} simulation package (VASP)~\cite{vasp1,vasp2}. Maximally localized Wannier functions (MLWFs) were constructed with the Wannier90~\cite{wannier90} to interpolate the electronic structure, which was combined with phonons computed in QE and processed using PERTURBO~\cite{Perturbo} with the corrected dielectric constants from HSE06. Finally, the Boltzmann transport equation (BTE) was solved within the relaxation-time approximation (RTA) to obtain phonon-limited electronic transport properties. The transport effective mass~\cite{gibbs2017effective} was evaluated from the Boltzmann-transport conductivity and carrier concentration obtained with BoltzTraP2~\cite{BoltzTraP2} within the constant-relaxation-time approximation (CRTA), according to $m^{*}(T,\mu) = n(T,\mu)\,e^{2}\tau/\sigma(T,\mu)$, where $e$ is the elementary charge, $\tau$ is the relaxation time, $n(T,\mu)$ is the carrier concentration at temperature $T$ and chemical potential $\mu$, and $\sigma(T,\mu)$ is the electrical conductivity. The underlying electronic structure was computed with VASP~\cite{vasp1,vasp2} using the PBE generalized-gradient approximation~\cite{PBE} and a plane-wave cutoff of 520~eV. All self-consistent calculations employed an energy convergence threshold of $10^{-8}$~eV. Projectors were evaluated in reciprocal space (LREAL = \texttt{.FALSE.}) to ensure accurate projections and velocities. $\Gamma$-centered $k$-point meshes were chosen sufficiently dense to converge Brillouin-zone integrals and transport coefficients. The crystal orbital Hamilton population (COHP) was computed with LOBSTER~\cite{LOBSTER}, and crystal structures were visualized with VESTA~\cite{vesta}.	
	
	Second-order interatomic force constants (IFCs) were computed using the finite-displacement method as implemented in PHONOPY~\cite{phonopy}. Third- and fourth-order IFCs were obtained via compressive sensing lattice dynamics (CSLD)~\cite{CSLD}. All DFT forces were evaluated in VASP~\cite{vasp1,vasp2} using the PBEsol functional~\cite{pbesol.PhysRevLett.100.136406}, a plane-wave cutoff of 520~eV, a $4\times4\times4$ supercell (512 atoms), and $\Gamma$-point $k$-mesh sampling. Finite-temperature phonon renormalization was treated within self-consistent phonon (SCPH) theory~\cite{scph1,scph2,scph3}. The $\kappa_{\rm L}$ was obtained by solving the Peierls--Boltzmann transport equation for phonons on uniform $q$-point meshes: $20\times20\times20$ for the harmonic-approximation plus three-phonon scattering (HA+3ph, $\kappa^{\rm HA}_{\rm 3ph}$) and the SCPH-renormalized three-phonon case (SCPH+3ph, $\kappa^{\rm SCPH}_{\rm 3ph}$), and $12\times12\times12$ when including both three- and four-phonon processes (SCPH+3ph+4ph, $\kappa^{\rm SCPH}_{\rm 3,4ph}$).

	\begin{figure*}[th!]
		\centering
		\includegraphics[width=1.0\linewidth]{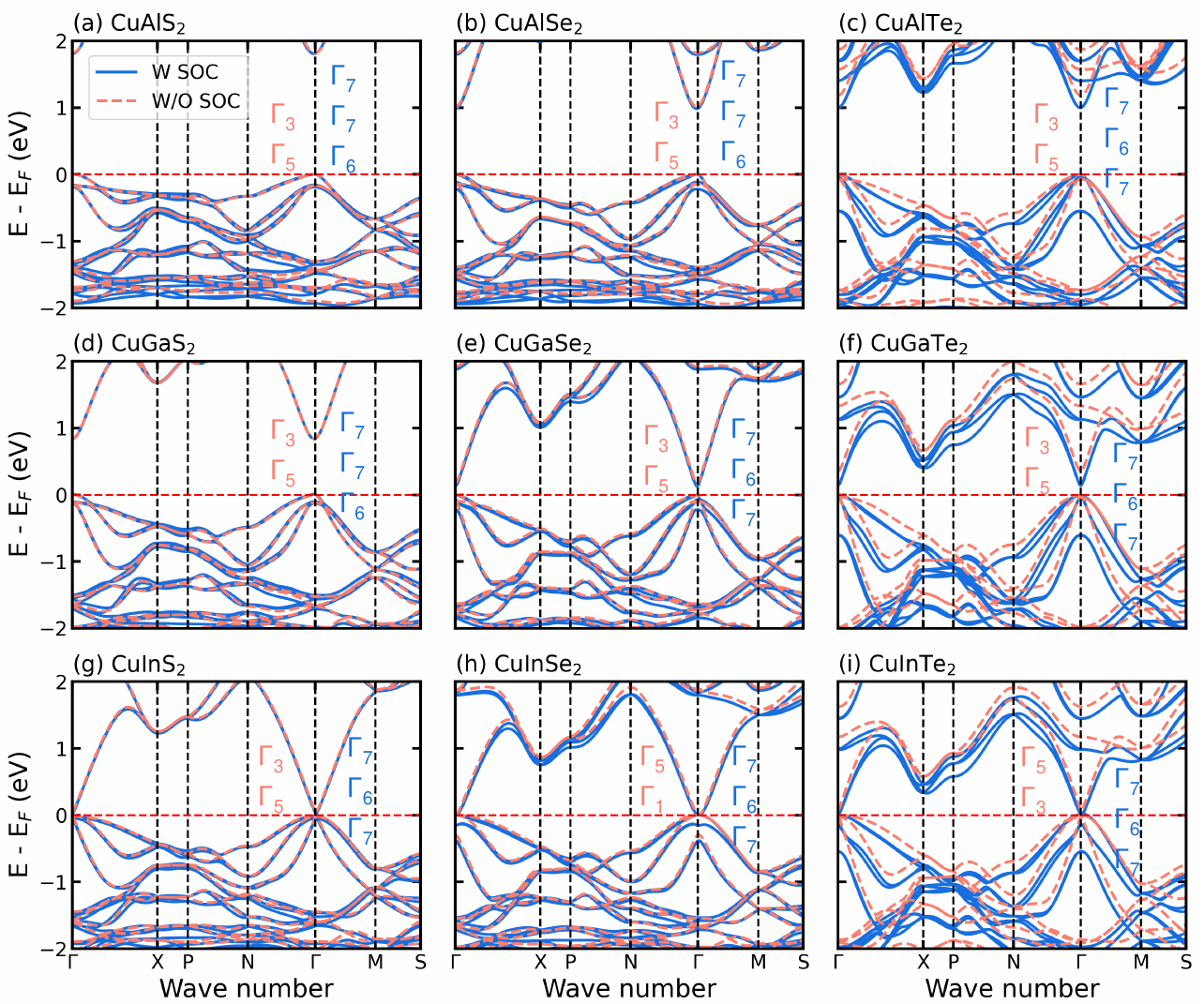}
		\caption{The electronic structures and irreducible representations of the top three valence bands at the $\Gamma$ point of Cu$MX_2$ ($M$ = Al, Ga, and In; $X$ = S, Se, and Te). (a)-(i) correspond to the compounds CuAlS$_2$, CuAlSe$_2$, CuAlTe$_2$, CuGaS$_2$, CuGaSe$_2$, CuGaTe$_2$, CuInS$_2$, CuInSe$_2$, and CuInTe$_2$, respectively. The blue and pink colors represent the electronic structures and irreducible representations with (W SOC) and without (W/O SOC) consideration SOC, respectively. Notably, $\Gamma_3$ ($\Gamma_1$) represents a non-degenerate state, while $\Gamma_5$ corresponds to a doubly degenerate state.}
		\label{band}
	\end{figure*}

	\section{RESULTS AND DISCUSSION}
	\subsection{Crystal structure and tetragonal distortion.} 
    The chalcopyrite structure is derived from the zinc blende (ZnS-type) structure, where two distinct cations (here, Cu and $M$) alternately occupy the cation sites of zinc blende, while the anion ($X$) resides in the tetrahedral interstices formed by the Cu and $M$ cations. In this arrangement, both Cu and $M$ cations are tetrahedrally coordinated by four $X$ anions, whereas each $X$ anion forms a tetrahedral coordination environment with two Cu and two $M$ cations, as illustrated in Fig.~\ref{crystal structure}(a). The nonequivalence between the Cu$^{+}$ and $M^{3+}$ cations reduces the symmetry from cubic (zinc blende) to tetragonal (chalcopyrite), resulting in slightly distorted tetrahedra. This structural deviation is quantified by the parameter $\eta = c/2a$~\cite{PhysRevB.29.1882}, where a value of $\eta \neq 1$ indicates tetrahedral distortion, with larger deviations corresponding to greater distortion. As listed in Table~\ref{lattice constant}, our calculated lattice constants for the nine Cu-based chalcopyrites agree well with experimental values~\cite{1953}. Additionally, our structural relaxations reveal that spin-orbit coupling (SOC) has a negligible effect on structure parameters. Notably, both theoretical and experimental $\eta$ values of these compounds exhibit slight deviations from unity. Specifically, CuAl$X_2$ and CuGa$X_2$ ($X$ = S, Se, and Te) display $\eta < 1$, whereas CuIn$X_2$ ($X$ = S, Se, and Te) exhibits $\eta > 1$. Previous studies have demonstrated that the tetragonal distortion significantly influences the electronic and TE properties of these compounds~\cite{https://doi.org/10.1002/adma.201400058,CuAg2024}. The distortion modifies crystal-field splitting, thereby altering the symmetry and state distribution of the electronic structure, which in turn modulates electronic transport properties~\cite{2008Complex}. Furthermore, it affects phonon dispersion relations and $\kappa_{\rm L}$ by modifying vibrational modes and phonon scattering rates ($1/\tau$)~\cite{zhao2014ultralow}. Given the critical role of these structural features in optimizing TE performance, a thorough understanding of their impact is essential for rational material design aimed at enhancing TE efficiency.

	\begin{table*}
		\scriptsize
		\setlength{\tabcolsep}{2pt} 
		\renewcommand{\arraystretch}{1.3} 
		\centering
		\caption{The calculated band gaps ($E_\mathrm{g}$), hole mobilities ($\mu_{\parallel}$, $\mu_{\perp}$, where $\parallel$ and $\perp$ indicate the directions parallel and perpendicular to the $c$ axis, respectively), electrical conductivities ($\sigma_{\parallel}$, $\sigma_{\perp}$), Seebeck coefficients ($S_{\parallel}$, $S_{\perp}$), high-frequency dielectric constants ($\epsilon^{\infty}_{\parallel}$, $\epsilon^{\infty}_{\perp}$), and static dielectric constants ($\epsilon^{0}_{\parallel}$, $\epsilon^{0}_{\perp}$) at 300~K for Cu$MX_2$ ($M$ = Al, Ga, and In; $X$ = S, Se, and Te). Unless otherwise noted, all transport quantities are evaluated at $n_\mathrm{h}=10^{18}\,\mathrm{cm}^{-3}$, except for CuInTe$_2$, $n_\mathrm{h}=6\times10^{18}\,\mathrm{cm}^{-3}$ is used to facilitate comparison with the experimental value $n_\mathrm{h}=5.75\times10^{18}\,\mathrm{cm}^{-3}$. Experimental values, where available, are given in parentheses.}
		\begin{tabular}{ccccccccccc}
			\hline
			\hline
			Comp.      & E$_{\mathrm{g}}^\mathrm{hse}$(eV) & $\mu^{\parallel}$,$\mu^{\perp}$(cm$^2$/Vs) & $\sigma^{\parallel}$,$\sigma^{\perp}$(S/m) & $S^{\parallel}$,$S^{\perp}$($\mu$V/K) & $\epsilon^{\infty}_{\parallel}$ & $\epsilon^{\infty}_{\perp}$ & $\epsilon^{0}_{\parallel}$ & $\epsilon^{0}_{\perp}$ & $\frac{1}{\epsilon^{\infty}_{\parallel}}-\frac{1}{\epsilon^0_{\parallel}}$ & $\frac{1}{\epsilon^{\infty}_{\perp}}-\frac{1}{\epsilon^0_{\perp}}$ \\
			\hline
			CuAlS$_2$  & 3.36 (3.46)$^a$ & 22,13             & 349,321              & 494,521           & 5.4 (4.8)$^j$ & 5.4 (4.9)$^j$  & 7.2 (7.1)$^j$    & 7.6 (8.1)$^j$   & 0.046 & 0.054 \\
			CuAlSe$_2$ & 2.68 (2.67)$^b$ & 47,25             & 764,399              & 441,493           & 5.9 (6.7)$^j$ & 6.0 (6.0)$^j$  & 7.6 (5.3)$^j$    & 8.1 (8.3)$^j$   & 0.038 & 0.043 \\
			CuAlTe$_2$ & 2.49 (2.06)$^c$ & 154,112           & 2484,1805            & 365,406           & 7.4           & 7.4            & 8.7              & 9.1             & 0.020 & 0.025 \\
			CuGaS$_2$  & 2.45 (2.43)$^d$ & 41,23             & 661,374              & 437,494           & 6.2 (6.1)$^j$ & 6.2 (6.2)$^j$  & 8.2 (7.6)$^j$    & 9.0 (8.9)$^j$   & 0.039 & 0.050 \\
			CuGaSe$_2$ & 1.72 (1.68)$^a$ & 52,26             & 837,427              & 376,454           & 7.2           & 7.1            & 9.3              & 10.0            & 0.031 & 0.041 \\
			CuGaTe$_2$ & 1.68 (1.23)$^e$ & 115,112 (112)$^h$ & 2034,1982 (2000)$^h$ & 380,404 (380)$^h$ & 8.8           & 8.5            & 10.3             & 10.7            & 0.017 & 0.024 \\
			CuInS$_2$  & 1.63 (1.53)$^a$ & 30,30             & 486,488              & 560,556           & 6.1           & 6.1            & 11.1             & 12.6            & 0.074 & 0.085 \\
			CuInSe$_2$ & 1.67 (1.01)$^f$ & 38,44             & 610,709              & 523,530           & 6.4 (8.5)$^j$ & 6.2 (9.5)$^j$  & 8.9 (15.2)$^j$   & 9.5 (16.0)$^j$  & 0.044 & 0.056 \\
			CuInTe$_2$ & 1.32 (0.96)$^g$ & 95,95 (90)$^i$    & 9121,9151 (8242)$^i$ & 253,273 (273)$^i$ & 8.6 (8.7)$^j$ & 8.5 (11.0)$^j$ & 10.9 (10.5)$^j$  & 11.5 (12.9)$^j$ & 0.024 & 0.031 \\
			\hline
			$^a$Ref.~\cite{gap1} &&&&&&&&&& \\
			$^b$Ref.~\cite{gap3} &&&&&&&&&& \\
			$^c$Ref.~\cite{gap5} &&&&&&&&&& \\
			$^d$Ref.~\cite{gap6} &&&&&&&&&& \\
			$^e$Ref.~\cite{gap2} &&&&&&&&&& \\
			$^f$Ref.~\cite{gap7} &&&&&&&&&& \\
			$^g$Ref.~\cite{gap4} &&&&&&&&&& \\
			$^h$Ref.~\cite{CuGaTe2} &&&&&&&&&& \\
			$^i$Ref.~\cite{CuInTe2} &&&&&&&&&& \\
			$^j$Ref.~\cite{10.1007/978-3-642-18865-7}
		\end{tabular}
		\label{band gap}
	\end{table*}    

    \begin{figure}[th!]
	\centering
	\includegraphics[width=1.0\linewidth]{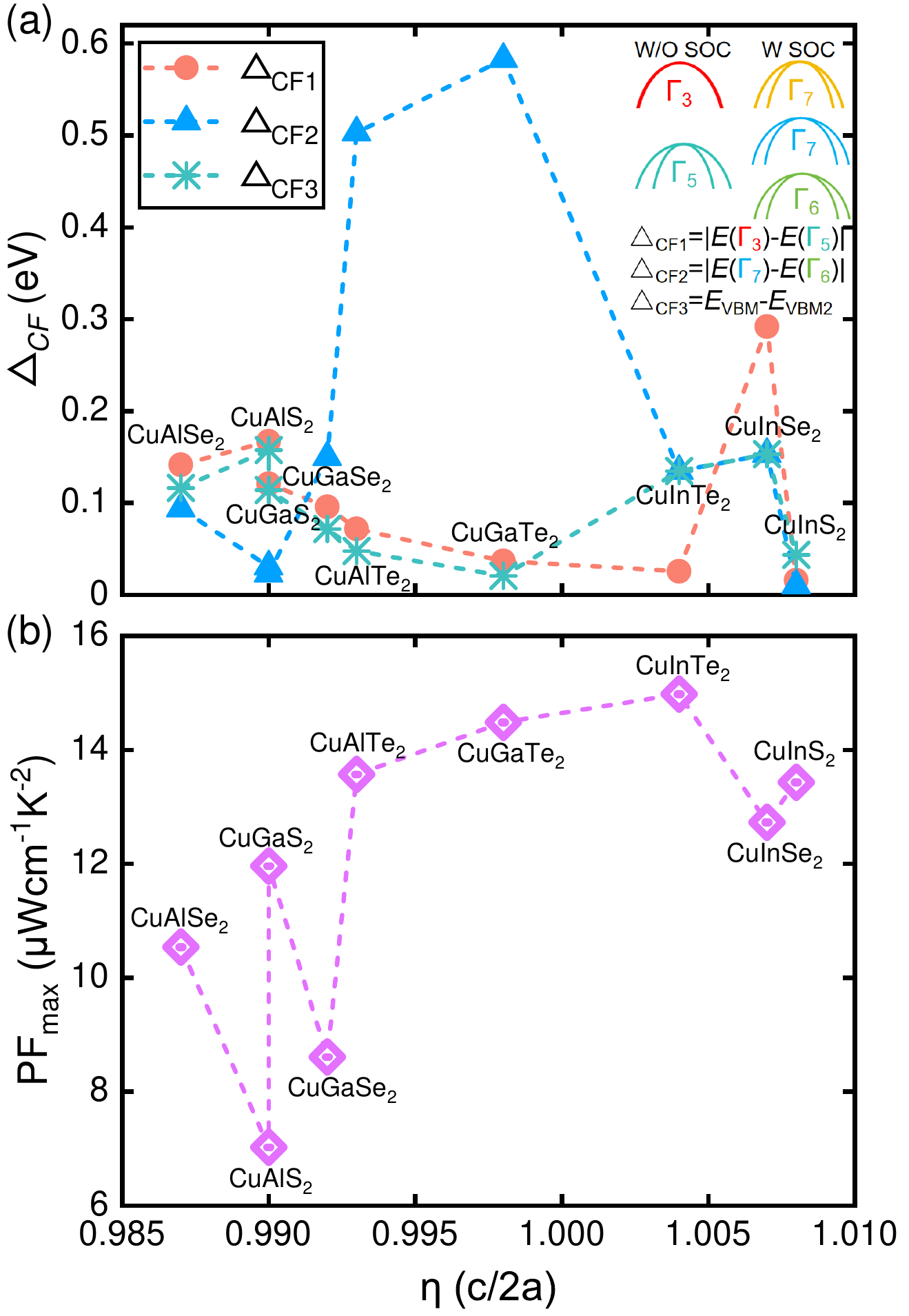}
	\caption{(a) and (b) show the dependence of the crystal-field splitting energy ($\Delta_{\mathrm{CF}}$) and the maximum power factor (PF$_\mathrm{max}$) of Cu$MX_2$ on the structural parameter $\eta$, respectively. The three measures of crystal-field splitting are defined as $\Delta_{\mathrm{CF1}} \equiv E(\Gamma_3)-E(\Gamma_5)$, $\Delta_{\mathrm{CF2}} \equiv E(\Gamma_7)-E(\Gamma_6)$, and $\Delta_{\mathrm{CF3}} \equiv E(\mathrm{VBM})-E(\mathrm{VBM2})$.}
	\label{crystalfield}
    \end{figure}

	\begin{figure*}[th!]
		\centering
		\includegraphics[width=1.0\linewidth]{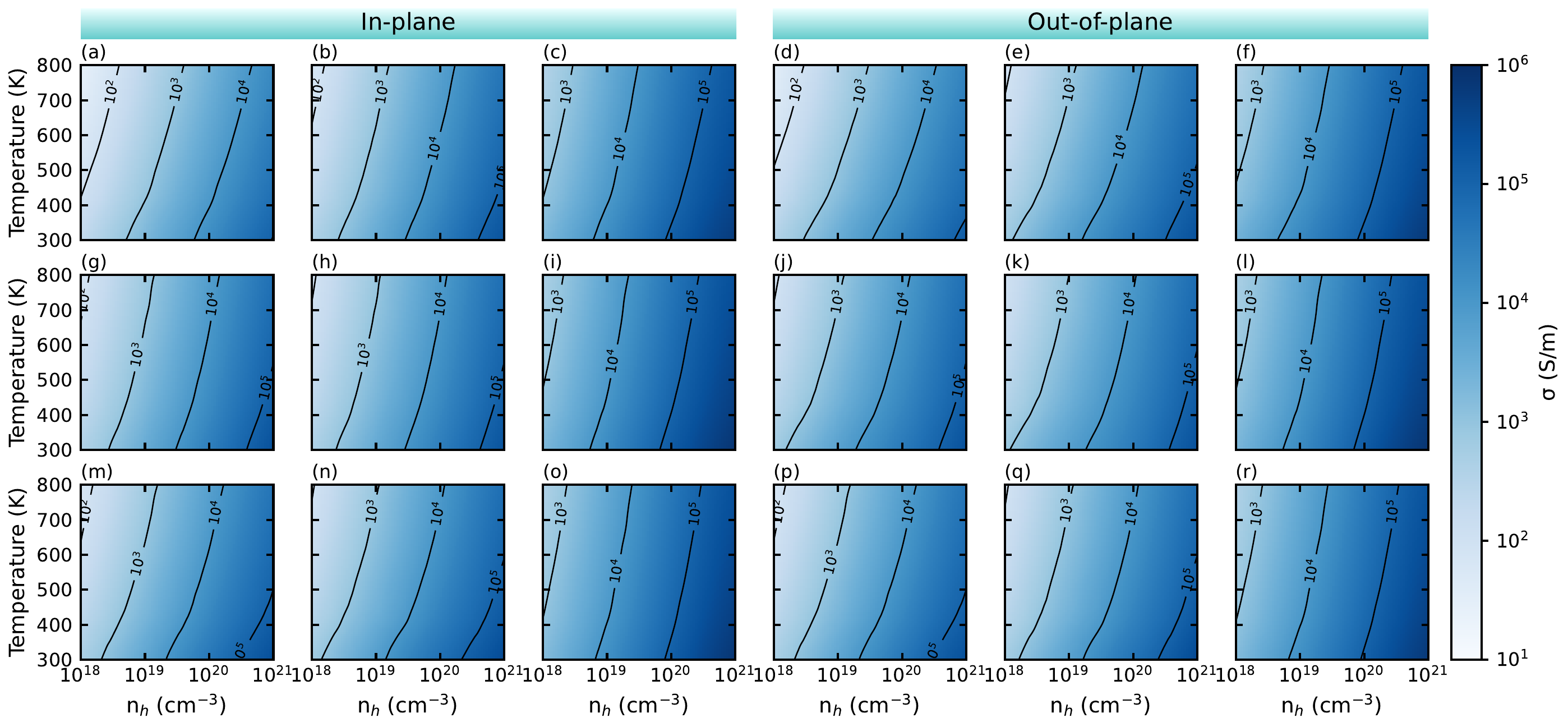}
		\caption{Electrical conductivity, $\sigma$, of Cu$MX_2$ ($M$ = Al, Ga, and In; $X$ = S, Se, and Te). (a)–(c), (g)–(i), (m)–(o) show the in-plane response ($\sigma^{\perp}$, within the $ab$ plane) for CuAlS$_2$, CuAlSe$_2$, CuAlTe$_2$, CuGaS$_2$, CuGaSe$_2$, CuGaTe$_2$, CuInS$_2$, CuInSe$_2$, and CuInTe$_2$, respectively. (d)–(f), (j)–(l), (p)–(r) show the out-of-plane response ($\sigma^{\parallel}$, along the $c$ axis) for the same sequence of compounds.}
		\label{sigma}
	\end{figure*}

	\subsection{Electronic structures}
	The orbital-projected electronic band structures, PDOS, and crystal orbital Hamilton population (COHP) for these materials calculated with PBE without SOC are depicted in Fig.~\ref{bandnosoc}. For clarity, the negative COHP (-COHP) is presented, where positive and negative values correspond to bonding and antibonding states, respectively. The band structures reveal that all nine compounds exhibit direct band gaps, with both valence band maxima (VBM) and conduction band minima (CBM) located at the Brillouin zone center ($\Gamma$ point). This degeneracy arises from symmetry-allowed $p$-$d$ coupling enabled by shared band representation elements~\cite{xiong2025forbidden}. Among these, CuAlS$_2$, CuAlSe$_2$, CuAlTe$_2$, and CuGaS$_2$ display relatively large band gaps, whereas CuGaSe$_2$ and CuGaTe$_2$ exhibit smaller band gaps. Notably, the CuIn$X_2$ ($X$ = S, Se, and Te) compounds show nearly vanishing band gaps at the PBE level, consistent with the systematic band gap underestimation characteristic of semilocal functionals. To address this limitation, we calculated band gaps using the screened hybrid functional HSE06~\cite{hse.10.1063/1.1564060}, with comparative experimental values provided in Table~\ref{band gap}. The HSE06-calculated band gaps demonstrate significantly improved agreement with experimental data. For CuAlS$_2$, CuAlSe$_2$, CuGaS$_2$, CuGaSe$_2$, and CuInS$_2$, our calculations yield results within 0.1~eV of experimental values. Conversely, for CuAlTe$_2$, CuGaTe$_2$, CuInSe$_2$, and CuInTe$_2$, the calculated band gaps underestimate experimental values by 0.43~eV, 0.34~eV, 0.66~eV, and 0.45~eV, respectively. This discrepancy likely stems from SOC exclusion in our HSE06 calculations, due to the heavy computational cost. We estimate SOC-induced band gap corrections using PBE functional, finding reductions of 0.18~eV and 0.22~eV for CuAlTe$_2$ and CuGaTe$_2$, respectively.

    Both the orbital-projected band structures and the PDOS reveal that the prominent peaks near the conduction band minimum (CBM) are associated with the $X$-$p$ and $M$-$p$ orbitals. The valence band, conversely, is dominated by Cu$^+$-$d$ and $X^{2-}$-$p$ orbital contributions due to strong $d$-$p$ coupling in the Cu-$X$ bonds~\cite{PhysRevB.37.8958,10.1063/1.4865257,https://doi.org/10.1002/adfm.202108532}. Conversely, the valence band is predominantly influenced by the Cu$^{+}$-$d$ and $X^{2-}$-$p$ orbitals, which is attributed to the strong $d$-$p$ coupling between Cu and $X$~\cite{PhysRevB.37.8958,10.1063/1.4865257,https://doi.org/10.1002/adfm.202108532}. As $X$ transitions from S to Se and subsequently to Te, its contribution to the valence band maximum (VBM) increases significantly. This trend can be explained by the fact that the valence band is primarily composed of the anti-bonding states of the Cu-$X$ bond, and the energy difference between the Cu-3$d$ and $X$-$p$ orbitals diminishes as one moves from S to Te. For instance, considering the compound Cu$M$S$_2$ (where $M$ represents Al, Ga, or In), it is observed that the energy of the Cu$^{+}$ 3$d$ orbitals exceeds that of the S$^{2-}$ 3$p$ orbitals~\cite{doi:10.1021/acs.chemrev.9b00600}. Consequently, the bonding and anti-bonding states primarily consist of S$^{2-}$-$p$ and Cu$^{+}$-$d$ orbitals, respectively, in the formation of the Cu-S chemical bond. As $X$ varies from S to Te, the atomic orbital energy difference decreases~\cite{doi:10.1021/acs.chemrev.9b00600}, leading to increasingly comparable contributions from Cu-$d$ and $X$-$p$ orbitals to both the bonding and anti-bonding states, as illustrated in Fig.~\ref{bandnosoc}.

    Finally, we investigate the influence of SOC on the electronic structures of Cu$MX_2$ ($M$ = Al, Ga, and In; $X$ = S, Se, and Te) compounds. As shown in Fig.~\ref{band}, SOC has a negligible effect on CuAlS$_2$, CuAlSe$_2$, CuGaS$_2$, and CuInS$_2$, but significantly modifies the band structures of CuAlTe$_2$, CuGaSe$_2$, CuGaTe$_2$, CuInSe$_2$, and CuInTe$_2$. The most pronounced changes occur near the VBM, which has a large contribution from $X$-$p$ orbitals. The doubly degenerate $\Gamma_5$ states (in the absence of SOC) split into $\Gamma_6$ and $\Gamma_7$ upon inclusion of SOC, whereas the singly degenerate $\Gamma_3$ state (labeled as $\Gamma_4$ in Ref.~\onlinecite{https://doi.org/10.1002/adma.201400058}) transforms into $\Gamma_7$ according to double group notation. As previously reported~\cite{https://doi.org/10.1002/adma.201400058,CuAg2024}, the magnitude of the crystal-field splitting energy ($\Delta_{\mathrm{CF}}$) plays a critical role in determining the PF of hole-doped systems, as it directly affects valence-band degeneracy. Therefore, accurate electron transport calculations must include SOC. Without SOC (Fig.~\ref{crystalfield}), the crystal-field splitting between $\Gamma_3$ and $\Gamma_5$ (denoted $\Delta_{\mathrm{CF1}}$) exhibits a dependence on the structural parameter $\eta$ consistent with previous findings~\cite{https://doi.org/10.1002/adma.201400058}, with $\Delta_{\mathrm{CF1}}$ reaching a minimum at $\eta \approx 1.0$. When SOC is included, the splitting between the $\Gamma_7$ and $\Gamma_6$ states derived from the $\Gamma_5$ manifold (denoted $\Delta_{\mathrm{CF2}}$) varies strongly with $\eta$ and reaches its maximum near $\eta \approx 1.0$, in stark contrast to the behavior of $\Delta_{\mathrm{CF1}}$. For CuAl$X_2$, CuGa$X_2$, and CuInS$_2$, where the doubly degenerate level $E(\Gamma_5)$ lies below the nondegenerate level $E(\Gamma_3)$ in the absence of SOC, the $\Gamma_5$-derived $\Gamma_6$/$\Gamma_7$ pair resides below the VBM. Consequently, $\Delta_{\mathrm{CF2}}$ does not affect the VBM degeneracy and is not an adequate descriptor of the relevant crystal-field splitting. Then, we define $\Delta_{\mathrm{CF3}}$ as the energy difference between the VBM and the next-lower valence state (VBM2) at the $\Gamma$ point, which is the pertinent measure for transport. Specifically, $\Delta_{\mathrm{CF3}}$ corresponds to the $\Gamma_6$–$\Gamma_7$ energy difference for CuInSe$_2$ and CuInTe$_2$, whereas for the remaining compounds it is the energy difference between the two $\Gamma_7$ levels. The evolution of $\Delta_{\mathrm{CF3}}$ with $\eta$ closely follows that of $\Delta_{\mathrm{CF1}}$, except for CuInTe$_2$ and CuInSe$_2$, where strong SOC raises the $\Gamma_5$-derived levels above the $\Gamma_3$-derived one (see Fig.~\ref{band}). Furthermore, $\Delta_{\mathrm{CF3}}$ is slightly smaller than $\Delta_{\mathrm{CF1}}$ for $\eta < 1.0$, whereas it becomes larger for $\eta > 1.0$.

	\subsection{Electronic Transport Properties}
	Because most Cu$MX_2$ ($M$ = Al, Ga, and In; $X$ = S, Se, and Te) compounds are $p$-type semiconductors~\cite{doi:10.1021/acs.chemrev.9b00600,CuInTe2,CuGaTe2}, we focus on $p$-type transport. A comparison between experiment and our calculations for CuGaTe$_2$ and CuInTe$_2$ at 300~K is summarized in Table~\ref{band gap}. For CuGaTe$_2$ at $n_\mathrm{h}=1\times10^{18}\,\mathrm{cm^{-3}}$, the calculated hole mobilities along the $a$ and $c$ axis are $\mu_{\perp}=112$ and $\mu_{\parallel}=115$~cm$^2$V$^{-1}$s$^{-1}$, the electrical conductivities are $\sigma_{\perp}=1982$ and $\sigma_{\parallel}=2034$~S\,m$^{-1}$, and the Seebeck coefficients are $S_{\perp}=404$ and $S_{\parallel}=380$~$\mu$V\,K$^{-1}$. These values closely match the measured $\mu=112$~cm$^2$V$^{-1}$s$^{-1}$, $\sigma=2000$~S\,m$^{-1}$, and $S=380$~$\mu$V\,K$^{-1}$ at the same $n_\mathrm{h}$ and $T$~\cite{CuGaTe2}. For CuInTe$_2$, evaluated at $n_\mathrm{h}=6.0\times10^{18}\,\mathrm{cm^{-3}}$ (to compare with experiment at $n_\mathrm{h}=5.75\times10^{18}\,\mathrm{cm^{-3}}$), we obtain $\mu_{\perp}=\mu_{\parallel}=95$~cm$^2$V$^{-1}$s$^{-1}$, $\sigma_{\perp}=9152$ and $\sigma_{\parallel}=9121$~S\,m$^{-1}$, and $S_{\perp}=273$ and $S_{\parallel}=253$~$\mu$V\,K$^{-1}$. These are in excellent agreement with the experimental values $\mu=90$~cm$^2$V$^{-1}$s$^{-1}$, $\sigma=8242$~S\,m$^{-1}$, and $S=273$~$\mu$V\,K$^{-1}$ at 300~K~\cite{CuInTe2}. The close agreement between calculated and measured $\mu$, $\sigma$, and $S$ indicates that electron--phonon scattering dominates charge transport in these materials and validates the accuracy and robustness of our computational approach across the Cu$MX_2$ series.

	\begin{figure*}[th!]
		\centering
		\includegraphics[width=1.0\linewidth]{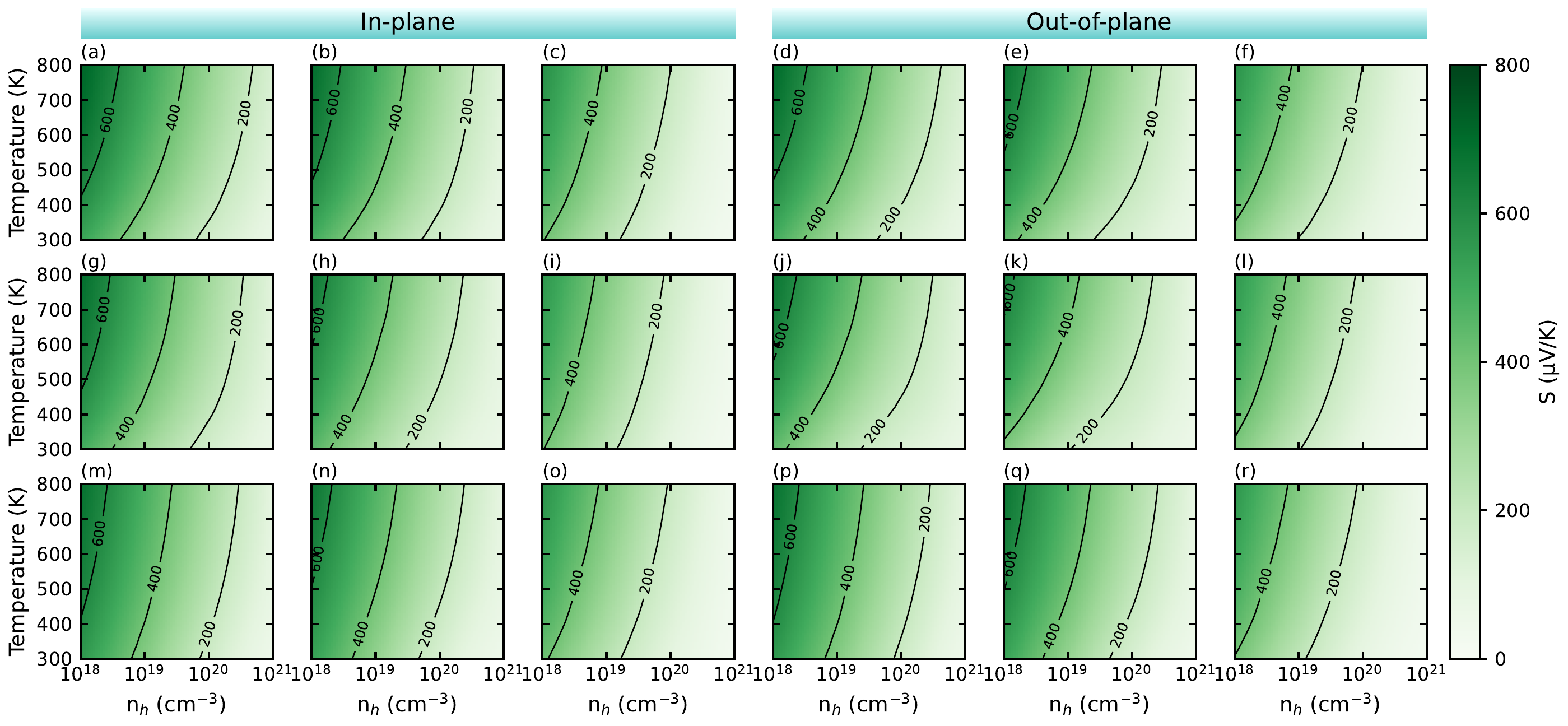}
		\caption{Seebeck coefficients, $S$, of Cu$MX_2$ ($M$ = Al, Ga, and In; $X$ = S, Se, and Te). (a)–(c), (g)–(i), (m)–(o) show the in-plane response ($S_{\perp}$, within the $ab$ plane) for CuAlS$_2$, CuAlSe$_2$, CuAlTe$_2$, CuGaS$_2$, CuGaSe$_2$, CuGaTe$_2$, CuInS$_2$, CuInSe$_2$, and CuInTe$_2$, respectively. (d)–(f), (j)–(l), (p)–(r) show the out-of-plane response ($S_{\parallel}$, along the $c$ axis) for the same sequence of compounds.}
		\label{seebeck}
	\end{figure*}

    Figure~\ref{sigma} shows the $\sigma$ of Cu$MX_2$ ($M$ = Al, Ga, and In; $X$ = S, Se, and Te) as a function of $n_{\mathrm{h}}$ and $T$. For all compounds, $\sigma$ increases monotonically with $n_{\mathrm{h}}$ and decreases with increasing $T$, as expected for semiconductors. For a fixed cation $M$, $\sigma$ follows the trend Cu$M$S$_2$ < Cu$M$Se$_2$ < Cu$M$Te$_2$ over nearly the entire $n_{\mathrm{h}}$–$T$ range. For example, at 300~K and $n_{\mathrm{h}}=1\times10^{18}$~cm$^{-3}$, we obtain $\sigma^{\perp}$ = 201, 399, and 1805~S\,m$^{-1}$ for CuAlS$_2$, CuAlSe$_2$, and CuAlTe$_2$, respectively. At fixed $X$, $\sigma$ also increases slightly from Al to Ga to In. Overall, the Cu$M$Te$_2$ series exhibits the highest $\sigma$. The conductivity anisotropy is generally weak, consistent with the small tetragonal distortion parameter $\eta$. When $\eta=1$ the structure is undistorted, and the anisotropy grows with $|\eta-1|$. Accordingly, CuAlS$_2$ and CuAlSe$_2$ display a modest but discernible anisotropy (Fig.~\ref{sigma}; $\eta=1.011$ and 1.013), whereas CuGaTe$_2$ and CuInTe$_2$ show negligible anisotropy because their $\eta$ values are very close to 1 (see Table~\ref{lattice constant}).

\begin{figure*}[th!]
	\centering
	\includegraphics[width=1.0\linewidth]{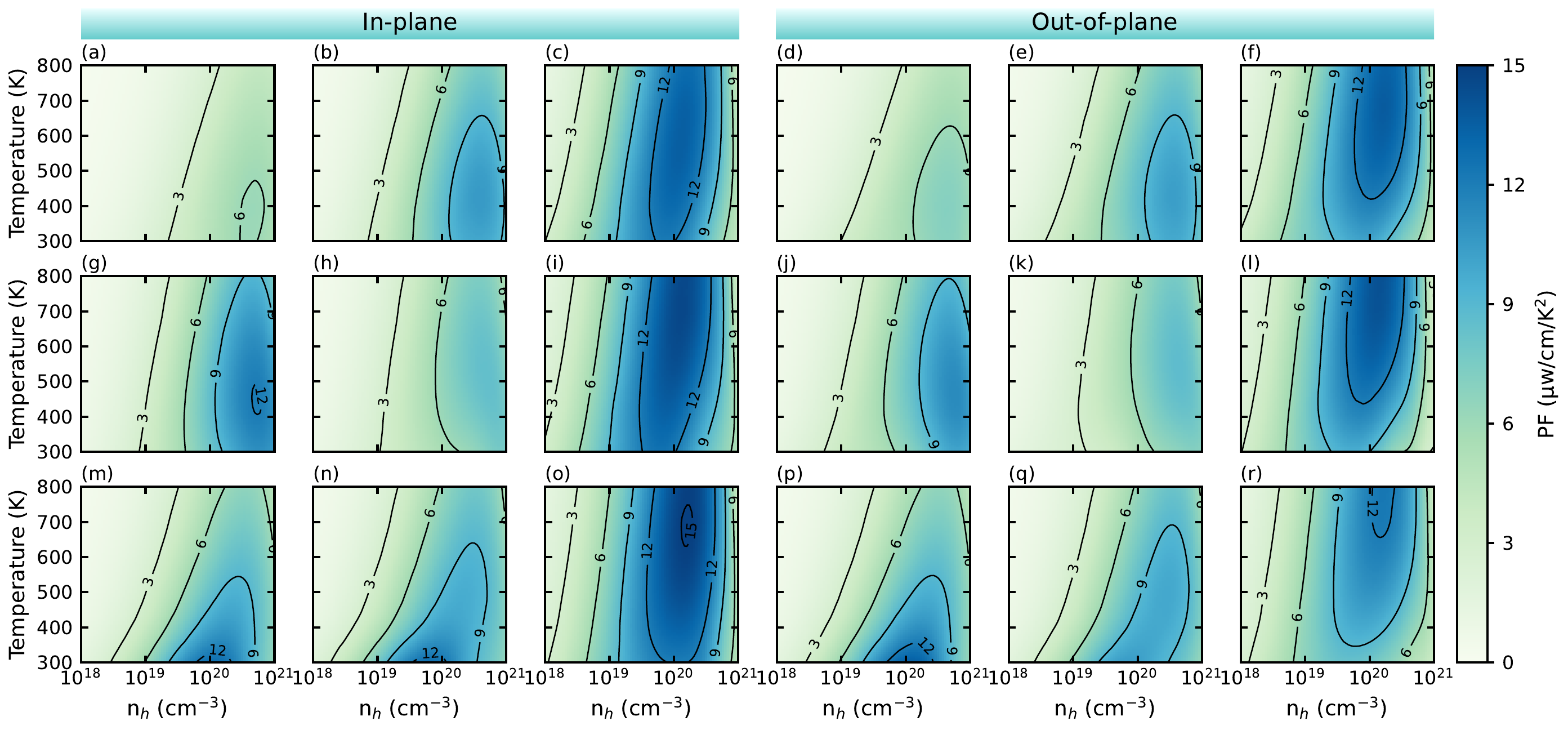}
	\caption{Power factor, PF, of Cu$MX_2$ ($M$ = Al, Ga, In; $X$ = S, Se, Te). (a)–(c), (g)–(i), (m)–(o) show the in-plane response (PF$^{\perp}$, within the $ab$ plane) for CuAlS$_2$, CuAlSe$_2$, CuAlTe$_2$, CuGaS$_2$, CuGaSe$_2$, CuGaTe$_2$, CuInS$_2$, CuInSe$_2$, and CuInTe$_2$, respectively. (d)–(f), (j)–(l), (p)–(r) show the out-of-plane response (PF$^{\parallel}$, along the $c$ axis) for the same sequence of compounds.}
	\label{pf}
\end{figure*}

	The observed trends in $\sigma$ can be rationalized in terms of the electron–phonon scattering rate ($1/\tau$) and the transport effective mass $m^*$, since for a fixed carrier concentration $\sigma = n_{\mathrm{h}} e^2 \tau / m^* \propto \tau/m^*$. Our calculated $1/\tau$ and $m^*$ are shown in Figs.~S3 and S4. Both $1/\tau$ and $m^*$ decrease along the series CuAlS$_2$ $\to$ CuGaS$_2$ $\to$ CuAlTe$_2$, consistent with the concomitant increase in $\sigma$. The scattering rates of CuInS$_2$ and CuInSe$_2$ are comparable, whereas $m^*$ is substantially larger in CuInS$_2$; hence the difference in $\sigma$ is predominantly governed by $m^*$. Within the tellurides Cu$M$Te$_2$ ($M$ = Al, Ga, and In) compounds, both $1/\tau$ and $m^*$ are very similar, leading to nearly indistinguishable $\sigma$. The mobility $\mu(T,n_{\mathrm{h}})$ is presented in Fig.~S5. At fixed $T$ and $n_{\mathrm{h}}$, $\mu$ increases markedly as $X$ is varied from S to Se to Te, explaining the ordering $\sigma(\mathrm{Te}) > \sigma(\mathrm{Se}) > \sigma(\mathrm{S})$. The enhanced $\mu$ in the tellurides arises from their smaller $m^*$ and lower $1/\tau$. As shown in Fig.~S4, heavier $X$ generally correlate with smaller $m^*$ in Cu$MX_2$ family. In polar semiconductors, electron scattering by polar optical phonons (POP) often dominates and is well captured by the Fr\"{o}hlich interaction~\cite{Frohlich01071954}. The long-wavelength electron and longitudinal optical (LO) phonon coupling matrix element is~\cite{Frohlich01071954,PhysRevLett.115.176401}
	\[
	g_{\mathbf{q}}
	= \frac{i}{|\mathbf{q}|}
	\left[
	\frac{e^2 \hbar \omega_{\mathrm{LO}}}{2 \varepsilon_0 N \Omega}
	\left(
	\frac{1}{\epsilon^{\infty}} - \frac{1}{\epsilon^{0}}
	\right)
	\right]^{\frac{1}{2}} ,
	\]
	\noindent where $\mathbf{q}$ is the phonon wave vector, $\Omega$ the unit-cell volume, $N$ the number of unit cells in the Born–von K\'{a}rm\'{a}n supercell, $e$ the elementary charge, $\varepsilon_0$ the vacuum permittivity, $\hbar$ the reduced Planck constant, $\epsilon^{\infty}$ and $\epsilon^{0}$ the high-frequency and static dielectric constants, respectively, and $\omega_{\mathrm{LO}}$ the frequency of the LO phonon. According to Table~\ref{band gap}, $\left(1/\epsilon^{\infty} - 1/\epsilon^{0}\right)$ decreases for both principal directions as $X$ becomes heavier (for fixed $M$). For a given $X$, the CuGa$X_2$ compounds exhibit smaller $\left(1/\epsilon^{\infty} - 1/\epsilon^{0}\right)$ than their Al- and In-based counterparts. Moreover, both CuGaTe$_2$ and CuGaTe$_2$ have particularly small $m^*$ (see Fig.~S4). These factors together account for their comparatively high $\sigma$.

    The $S$ of Cu$MX_2$ ($M$ = Al, Ga, and In; $X$ = S, Se, and Te) compounds are presented in Fig.~\ref{seebeck}. As expected, $S$ decreases with increasing $n_{\mathrm{h}}$, while it increases with rising temperature $T$, showing an inverse correlation with $\sigma$. This behavior indicates a strong interplay between $\sigma$ and $S$. The origin lies in the fact that $S$ is proportional to the density-of-states effective mass ($m_\mathrm{d}^*$), which, in the case of a single parabolic band, is directly proportional to the transport effective mass $m^{*}$. For instance, CuAlS$_2$ and CuInTe$_2$ possess the largest and smallest $m^{*}$, respectively, and therefore exhibit the highest and lowest $S$ at a given $n_{\mathrm{h}}$ and $T$. Compared with $\sigma$, the anisotropy of $S$ is even smaller in all these compounds, a trend that is commonly observed in semiconductors~\cite{https://doi.org/10.1002/adma.202104908,hzj}.

	The calculated PF of the Cu$MX_2$ compounds are presented in Fig.~\ref{pf}. Owing to the opposite dependence of $\sigma$ and $S$ on $n_{\mathrm{h}}$ and $T$, the PF for all these compounds initially increases and subsequently decreases with increasing $n_{\mathrm{h}}$. At 300~K, the three compounds exhibiting the highest PF values are CuInS$_2$ \big(PF = 13.4~$\mu$W\,cm$^{-1}$\,K$^{-2}$ along the $c$-axis at $n_{\mathrm{h}}$ = $1\times 10^{20}$~cm$^{-3}$\big), CuGaTe$_2$ \big(PF = 12.6~$\mu$W\,cm$^{-1}$\,K$^{-2}$ along the $a$-axis at $n_{\mathrm{h}}$ = $6\times 10^{19}$~cm$^{-3}$\big), and CuAlTe$_2$ \big(PF = 12.2~$\mu$W\,cm$^{-1}$\,K$^{-2}$ along the $a$-axis at $n_{\mathrm{h}}$ = $7\times 10^{19}$~cm$^{-3}$\big). At 800~K, the top three are CuInTe$_2$ \big(PF = 14.8~$\mu$W\,cm$^{-1}$\,K$^{-2}$ along the $a$-axis at $n_{\mathrm{h}}$ = $2\times 10^{20}$~cm$^{-3}$\big), CuGaTe$_2$ \big(PF = 14.3~$\mu$W\,cm$^{-1}$\,K$^{-2}$ along the $a$-axis at $n_{\mathrm{h}}$ = $2\times 10^{20}$~cm$^{-3}$\big), and CuAlTe$_2$ \big(PF = 13.4~$\mu$W\,cm$^{-1}$\,K$^{-2}$ along the $a$-axis at $n_{\mathrm{h}}$ = $2\times 10^{20}$~cm$^{-3}$\big). These results are in good agreement with experimental values. For instance, the measured PF of CuGaTe$_2$ is 15.0~$\mu$W\,cm$^{-1}$\,K$^{-2}$ at 850~K~\cite{CuGaTe2} and 14.0~$\mu$W\,cm$^{-1}$\,K$^{-2}$ at 794~K~\cite{2020Origin}, closely matching our calculated 14.3~$\mu$W\,cm$^{-1}$\,K$^{-2}$. Similarly, the experimental PF of CuInTe$_2$ is 13.0~$\mu$W\,cm$^{-1}$\,K$^{-2}$ at 850~K~\cite{CuInTe2}, consistent with our calculated value of 14.8~$\mu$W\,cm$^{-1}$\,K$^{-2}$. Finally, the variation of the maximum PF (PF$_{\mathrm{max}}$) with the structural parameter $\eta$ is shown in Fig.~\ref{crystalfield}(b). Compounds with $\eta \approx 1.0$ generally exhibit higher PF$_{\mathrm{max}}$ values compared to those deviating from unity. Notably, the largest PF$_{\mathrm{max}}$ occurs at $\eta = 1.03$ (CuInTe$_2$). Moreover, an oscillatory dependence of PF$_{\mathrm{max}}$ on $\eta$ is observed for $\eta < 0.995$, despite $\Delta_{\mathrm{CF3}}$ showing an almost monotonic increase from CuAlTe$_2$ to CuAlS$_2$. This non-monotonicity likely originates from the intricate competition between $\sigma$ and $S$.

	\begin{figure*}[th!]
		\centering
		\includegraphics[width=1.0\linewidth]{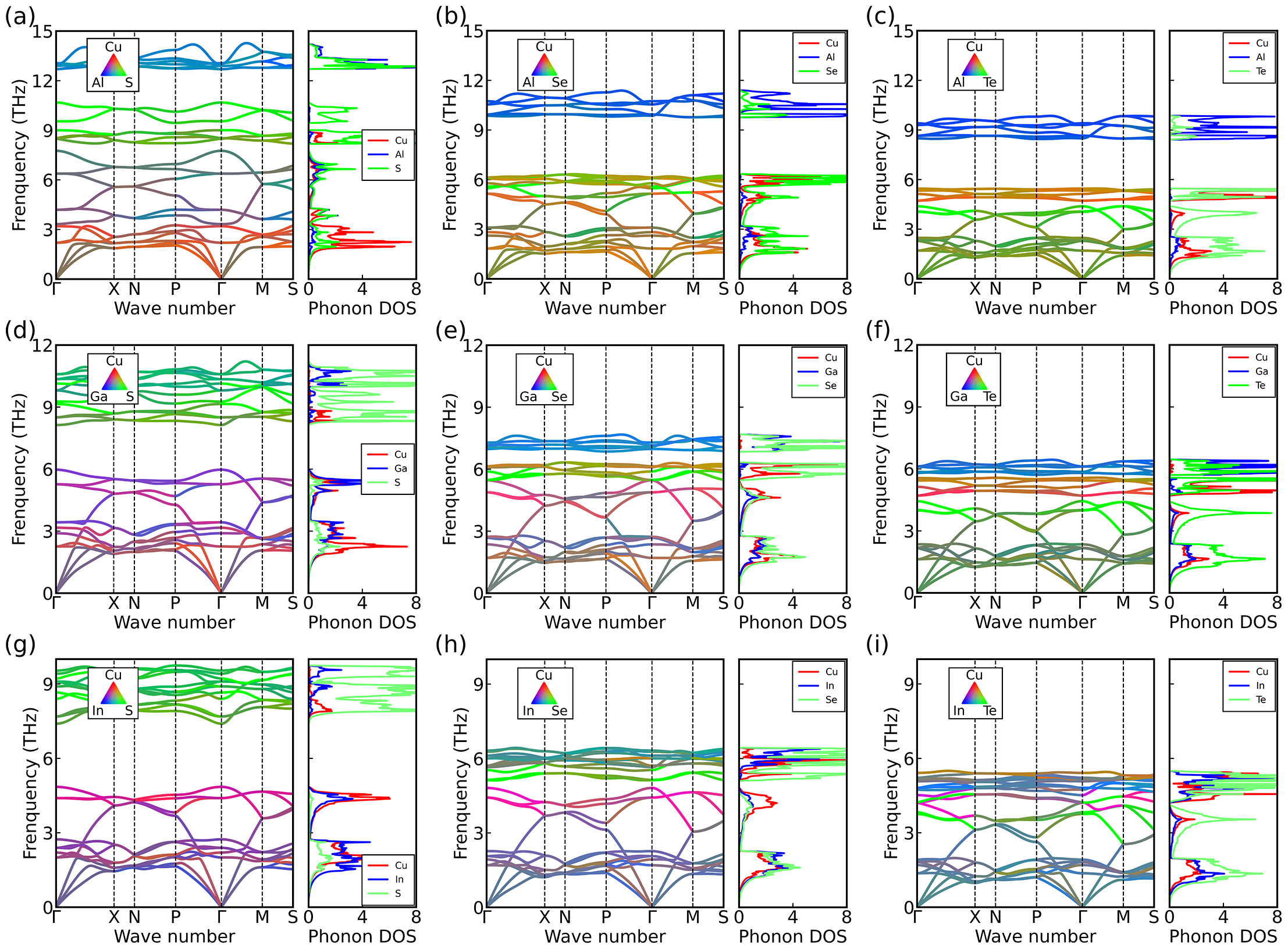}
		\caption{Phonon dispersion relations and phonon density of states (PhDOS) of Cu$MX_2$ ($M$ = Al, Ga, and In; $X$ = S, Se, and Te) calculated at 300 K. (a–i) correspond to CuAlS$_2$, CuAlSe$_2$, CuAlTe$_2$, CuGaS$_2$, CuGaSe$_2$, CuGaTe$_2$, CuInS$_2$, CuInSe$_2$, and CuInTe$_2$, respectively. The color scale indicates the atomic character of the phonon modes (atom-projected dispersions).}
		\label{phonon}
	\end{figure*}

	\begin{figure}[th!]
		\centering
		\includegraphics[width=1.0\linewidth]{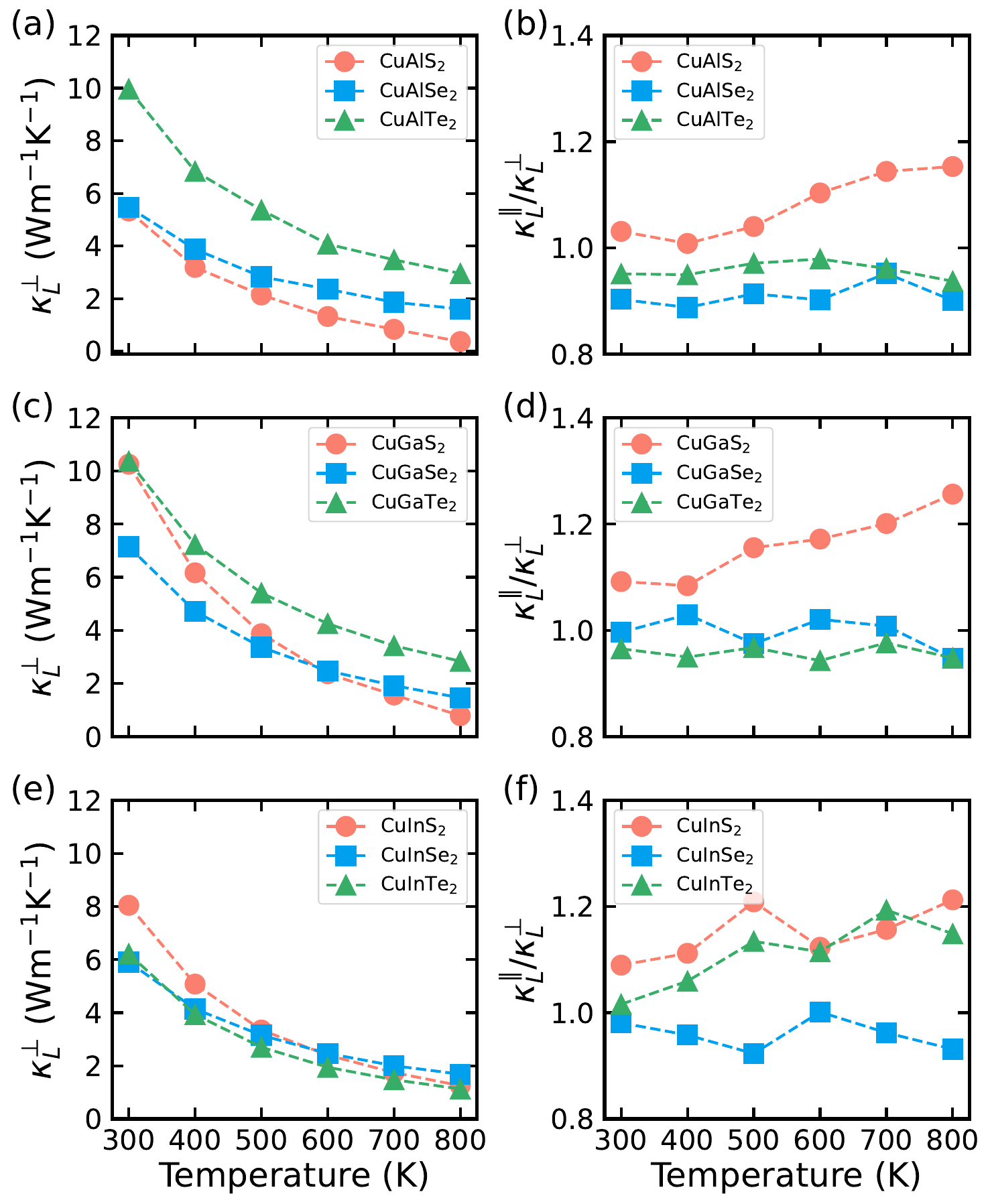}
		\caption{(a), (c), and (e) are the temperature dependence of $\kappa_{\mathrm{L}}^{\perp}$ (perpendicular to the $c$ axis) for CuAl$X_2$, CuGa$X_2$, and CuIn$X_2$ ($X$ = S, Se, and Te), respectively. (b), (d), and (f) are the corresponding ratios of $\kappa_{\mathrm{L}}^{\parallel}$ (along to the $c$ axis) to $\kappa_{\mathrm{L}}^{\perp}$ for the same set of compounds.}
		\label{kappa}
	\end{figure}

	\begin{figure*}[th!]
		\centering
		\includegraphics[width=1.0\linewidth]{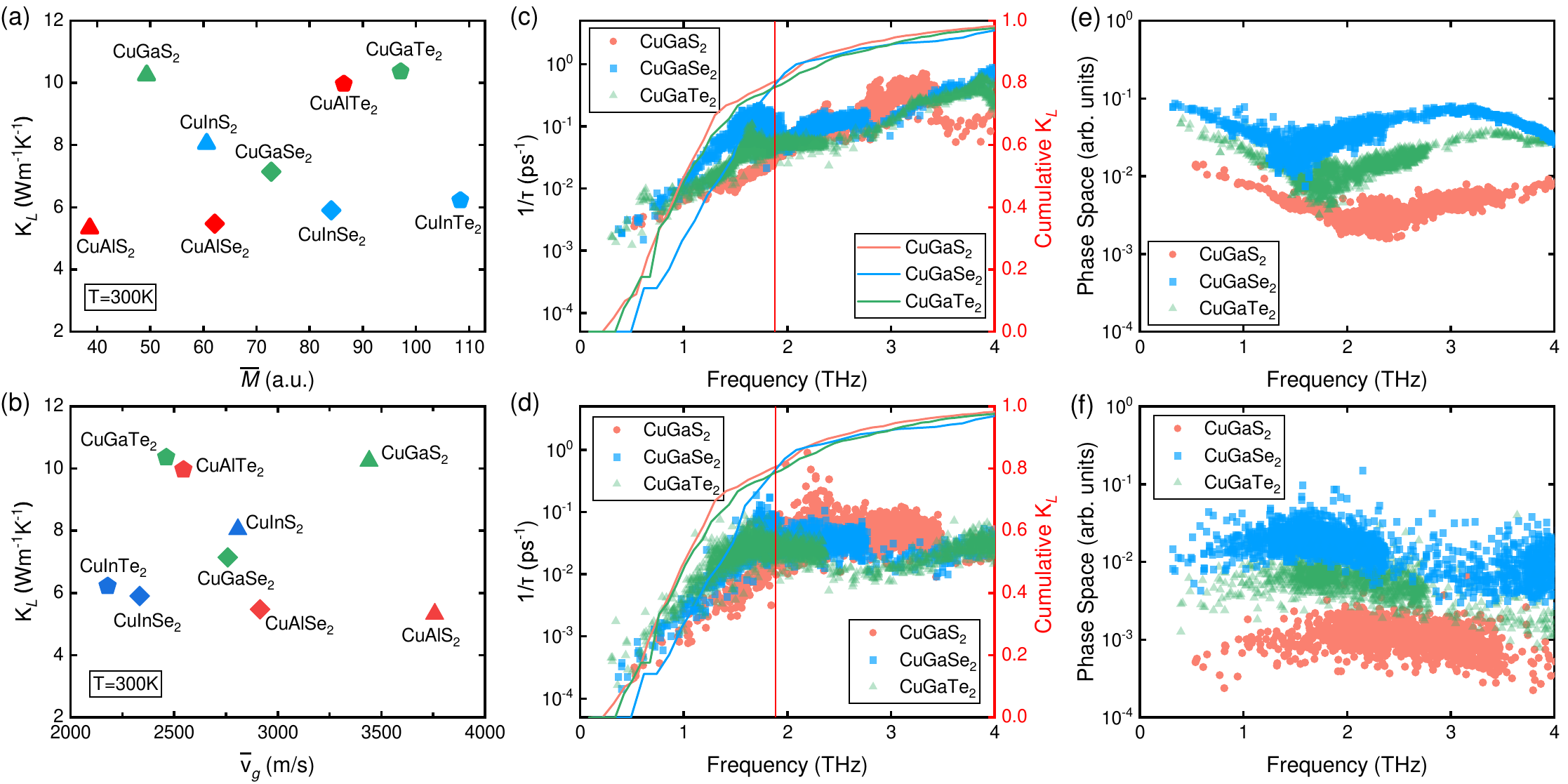}
		\caption{(a) and (b) depict the dependence of $\kappa_{\mathrm{L}}^{\perp}$ of Cu$MX_2$ ($M$ = Al, Ga, and In; $X$ = S, Se, and Te) on the phonon group velocity $\overline{\nu_{\mathrm{g}}}$ and the average atomic mass $\overline{M}$ at 300~K, respectively. (c) and (d) present the variation of the three- and four-phonon scattering rates, and the cumulative $\kappa_{\mathrm{L}}$ of CuGa$X_2$, as a function of phonon frequency. (e) and (f) show the dependence of the three- and four-phonon weighted phase space of CuGa$X_2$ on phonon frequency.}
		\label{ss}
	\end{figure*}

	\begin{figure*}[th!]
		\centering
		\includegraphics[width=1.0\linewidth]{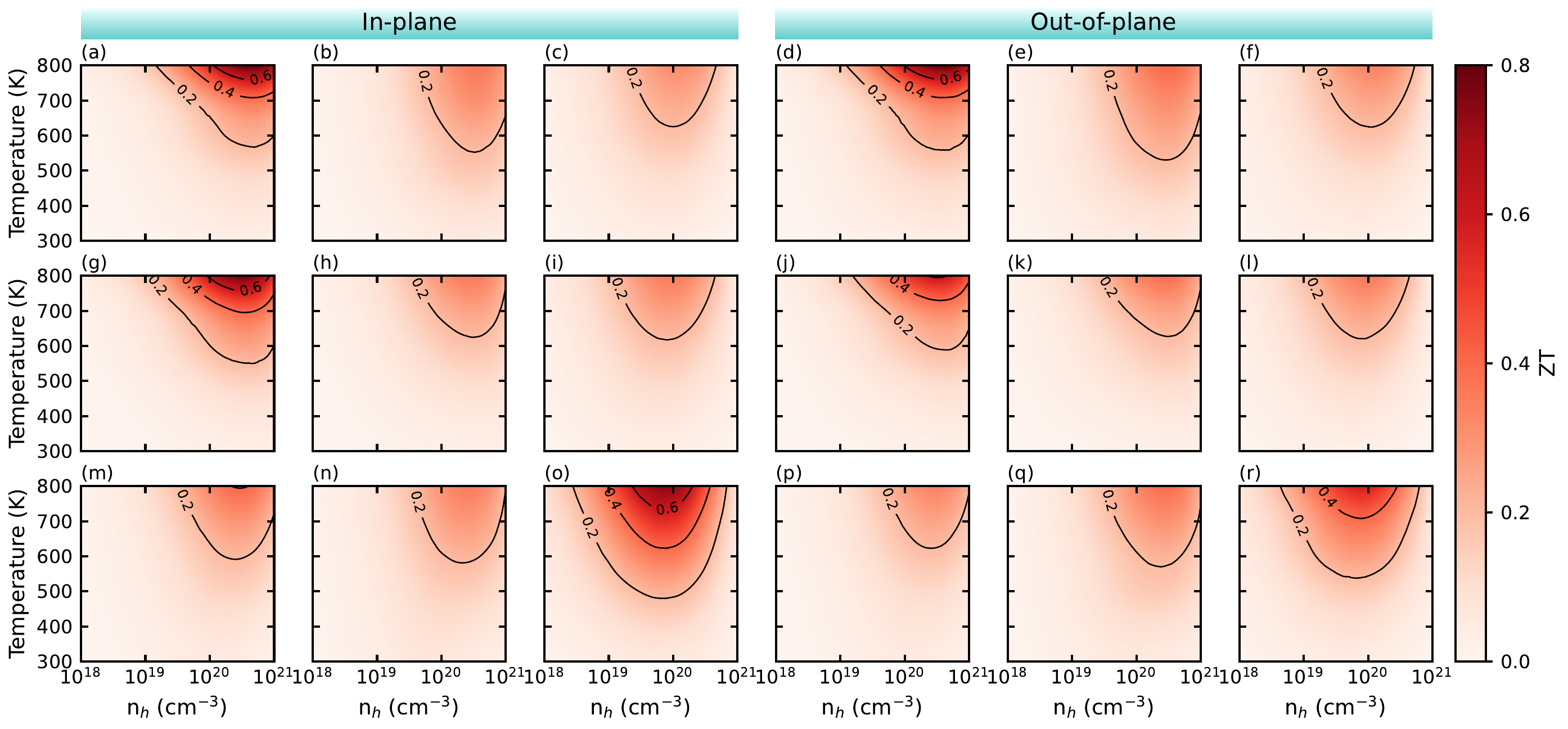}
		\caption{The thermoelectric figure of merit, $ZT$, of Cu$MX_2$ ($M$ = Al, Ga, and In; $X$ = S, Se, and Te). (a)–(c), (g)–(i), and (m)–(o) present the $ZT^{\perp}$ (in-plane) for CuAlS$_2$, CuAlSe$_2$, CuAlTe$_2$, CuGaS$_2$, CuGaSe$_2$, CuGaTe$_2$, CuInS$_2$, CuInSe$_2$, and CuInTe$_2$, respectively. (d)-(f), (j)-(l), and (p)-(r) show the corresponding $ZT^{\parallel}$ (out-of-plane) for CuAlS$_2$, CuAlSe$_2$, CuAlTe$_2$, CuGaS$_2$, CuGaSe$_2$, CuGaTe$_2$, CuInS$_2$, CuInSe$_2$, and CuInTe$_2$, respectively.}
		\label{zt}
	\end{figure*}

	\subsection{Phonon dispersion and lattice thermal conductivity}
	Our previous study demonstrated that the PBEsol exchange–correlation functional outperforms other functionals in calculating $\kappa_{\mathrm{L}}$ of rock-salt and zinc-blende semiconductors~\cite{PhysRevB.110.035205}. Therefore, the PBEsol functional is adopted in the present work for calculating $\kappa_{\mathrm{L}}$. The phonon dispersion relations and phonon density of states (PhDOS) of Cu$MX_2$ ($M$ = Al, Ga, and In; $X$ = S, Se, and Te) at 300~K, obtained using PBEsol, are shown in Fig.~\ref{phonon}. The primitive unit cell of each Cu$MX_2$ compound contains eight atoms, resulting in a total of 24 phonon branches at each wave vector: three acoustic and twenty-one optical modes. In Fig.~\ref{phonon}, the atomic contributions to the phonon bands from Cu, $M$, and $X$ atoms are color-coded in red, blue, and green, respectively. The phonon frequency $\omega$ scales proportionally with the bonding strength $k$ and inversely with the average atomicass $\overline{M}$, following the approximate relation $\omega \propto \sqrt{k/\overline{M}}$. Consequently, heavier atoms predominantly contribute to low-frequency phonon modes, whereas lighter atoms mainly appear in the high-frequency range. For instance, in CuAlS$_2$, the low-frequency modes are primarily associated with Cu atoms (63.55~a.u.) due to their higher atomic mass compared to Al (26.98~a.u.) and S (32.06~a.u.). However, as $X$ is varied from S to Se (78.97~a.u.) and Te (127.60~a.u.), Cu contribution to low-frequency phonons decreases markedly. Simultaneously, the maximum phonon frequency decreases sequentially in the order CuAlS$_2$ ($\sim$15~THz) $>$ CuAlSe$_2$ ($\sim$11~THz) $>$ CuAlTe$_2$ ($\sim$5~THz), with the frequency distribution becoming increasingly dense. This evolution is apparent in the PhDOS, where the phonon bands dominated by $X$ atoms progressively shift toward lower frequencies with increasing $X$ atomic mass. A similar trend is also observed when the atomic mass of $M$ increases.

	The calculated $\kappa_{\mathrm{L}}$, including both three-phonon (3ph) and four-phonon (4ph) scattering processes based on renormalized second-order force constants at each temperature, are presented in Fig.~\ref{kappa}. Here, $\kappa_{\mathrm{L}}^{\perp}$ and $\kappa_{\mathrm{L}}^{\parallel}$ denote the values along the $a$- and $c$-axes, respectively. Our calculated $\kappa_{\mathrm{L}}$ at 300~K are in good agreement with previous theoretical results~\cite{plata2022charting,https://doi.org/10.1002/adfm.202005861} and with experimental data for CuInTe$_2$ (6~W\,m$^{-1}$\,K$^{-1}$)~\cite{PhysRevB.105.245204} and CuGaTe$_2$ (10.7~W\,m$^{-1}$\,K$^{-1}$)~\cite{plata2022charting}. As in other zinc-blende-based compounds~\cite{PhysRevX.10.041029,PhysRevB.110.035205}, the contribution of 4ph scattering to $\kappa_{\mathrm{L}}$ is relatively small (see Fig.~S7), attributable to the low anharmonicity of these zinc-blende-like materials~\cite{PhysRevB.110.035205}. The $\kappa_{\mathrm{L}}^{\parallel} / \kappa_{\mathrm{L}}^{\perp}$ ratios fall within the range 0.8–1.2, indicating weak anisotropy, consistent with the nearly undistorted crystal structure ($\eta \approx 1$). Interestingly, the $\kappa_{\mathrm{L}}^{\parallel} / \kappa_{\mathrm{L}}^{\perp}$ ratio exhibits a temperature dependence. For Cu$M$S$_2$ ($M$ = Al, Ga, In) compounds, the ratio remains greater than 1 and increases with temperature. In contrast, for Cu$M$Se$_2$ and Cu$M$Te$_2$, the ratio shows a stronger dependence on the specific $M$ element.

    As reported previously~\cite{plata2022charting}, Cu$MX_2$ compounds exhibit an anomalous dependence of the $\kappa_{\mathrm{L}}$ on $X$. As shown in Fig.~\ref{ss}(a), for a fixed $M$ the Se-based compounds generally have lower $\kappa_{\mathrm{L}}$ than their Te-based counterparts. For example, CuGaSe$_2$ has a lower $\kappa_{\mathrm{L}}$ than CuGaTe$_2$, despite Te being heavier and less electronegative. To rationalize this behavior, we examine the dependence of $\kappa_{\mathrm{L}}$ at 300~K on the $\overline{\nu_g}$ for these compounds [Fig.~\ref{ss}(b)]. Although $\overline{\nu_g}$ is nearly linearly correlated with $\overline{M}$ (Fig.~S8), the dependence of $\kappa_{\mathrm{L}}$ on $\overline{M}$ is nontrivial and varies with $M$. Within kinetic theory~\cite{tritt2005thermal}, $\kappa_{\mathrm{L}} \propto \overline{\nu_g}^{\,2}\tau$, where $\tau$ is the phonon relaxation time ($\tau$ is the inverse of the phonon–phonon scattering rate). Deviations from the expected $\kappa_{\mathrm{L}} \propto \overline{\nu_g}^{\,2}$ trend therefore highlight the critical role of phonon–phonon scattering. Panels (c) and (d) of Fig.~\ref{ss} present the 3ph and 4ph scattering rates and the cumulative $\kappa_{\mathrm{L}}$ as functions of phonon frequency for CuGa$X_2$. The cumulative $\kappa_{\mathrm{L}}$ shows that phonons below 1.9~THz contribute about 80\% of the total $\kappa_{\mathrm{L}}$, underscoring the dominant roles of acoustic and low-frequency optical modes. In this frequency range, the 3ph scattering rates of CuGaSe$_2$ are substantially higher than those of CuGaTe$_2$ and CuGaS$_2$, i.e., they follow the inverse order of $\kappa_{\mathrm{L}}$, consistent with the observed anomalous $X$ dependence. The enhanced 3ph scattering in CuGaSe$_2$ is attributable to its larger weighted phase space. By contrast, below 1.0~THz (where acoustic modes dominate), the 4ph scattering rates are smaller than that of 3ph; between 1.0 and 1.9~THz, the 4ph scattering rates become comparable to 3ph rates. These results indicate that, for a given $M$, the anomalous $X$ dependence of $\kappa_{\mathrm{L}}$ arises primarily from strengthened low-frequency three-phonon scattering in the Se-based compounds.

	\subsection{Figure of merit $ZT$}
	With all TE transport properties in hand, we evaluate the dimensionless figure of merit $ZT$ for these compounds as functions of the $n_{\mathrm{h}}$ and $T$, as shown in Fig.~\ref{zt}. Because $ZT$ scales with $T$ and inversely with $\kappa_{\mathrm{L}}$, and because $\kappa_{\mathrm{L}}$ decreases with increasing $T$, the maximum $ZT$ for all compounds appears at $800\,\mathrm{K}$, which is the upper bound of our temperature range and a temperature at which several of these compounds are known to exhibit high TE performance~\cite{2020Origin,doi:10.1021/acsaem.0c01867}. For all materials, $ZT$ exhibits a characteristic rise and subsequent decline with increasing $n_{\mathrm{h}}$, with the peak occurring near $n_{\mathrm{h}}\sim 10^{20}\,\mathrm{cm^{-3}}$. Among the compounds studied, the largest $ZT$ is obtained for CuAlS$_2$ at $800\,\mathrm{K}$, with $ZT^{\perp}\approx ZT^{\parallel}=0.76$, owing primarily to its low $\kappa_{\mathrm{L}}$ at this temperature. The $ZT^{\perp}$ values of CuGaS$_2$ and CuInTe$_2$ are also comparatively large, at 0.74 and 0.71, respectively. Notably, our calculated $ZT^{\perp}$ for CuInTe$_2$ (0.71) is in very good agreement with the experimental value (0.74)~\cite{2020Origin}. Although CuGaTe$_2$ attains the highest PF among these compounds, $14.8\,\mu\mathrm{W\,cm^{-1}\,K^{-2}}$ at $800\,\mathrm{K}$, its relatively large $\kappa_{\mathrm{L}}$ ($2.84\,\mathrm{W\,m^{-1}\,K^{-1}}$ at $800\,\mathrm{K}$) suppresses its $ZT$. By contrast, $\kappa_{\mathrm{L}}$ for CuAlS$_2$ and CuGaS$_2$ at $800\,\mathrm{K}$ is $0.37\,\mathrm{W\,m^{-1}\,K^{-1}}$ and $0.79\,\mathrm{W\,m^{-1}\,K^{-1}}$, respectively, and the favorable balance between PF and $\kappa_{\mathrm{L}}$ yields the highest $ZT$ values in these two compounds. We note that our $\kappa_{\mathrm{L}}$ calculations assume perfect crystals and employ a rigid-band approximation for doping, therefore, phonon scattering from defects, grain boundaries, and related microstructural features is neglected. Such scattering mechanisms can substantially reduce $\kappa_{\mathrm{L}}$, especially under the heavy doping commonly used in experiments. Consequently, our calculated $ZT$ values likely represent lower bounds relative to experimental measurements.

	\section{CONCLUSIONS}
    In summary, we performed a comprehensive, first-principles investigation of the thermoelectric properties of copper-based chalcopyrites Cu$MX_2$ ($M=$ Al, Ga, and In; $X=$ S, Se, and Te). Electronic transport was computed by explicitly including electron–phonon coupling on top of PBE electronic structures with spin–orbit coupling and with corrected band gaps and dielectric constants using HSE06. Lattice thermal conductivities were obtained by solving the phonon Boltzmann transport equation (BTE) with three- and four-phonon scattering processes, using temperature-renormalized second-order force constants across the full temperature range considered. All compounds are direct band gap semiconductors with both the valence-band maximum and conduction-band minimum located at the $\Gamma$ point. Pronounced SOC-induced valence-band splittings that overlap with crystal-field splittings are found in CuAlTe$_2$, CuGaSe$_2$, CuGaTe$_2$, CuInSe$_2$, and CuInTe$_2$, which reduce the valence-band degeneracy and underscore the necessity of including SOC in compounds containing In and Te. For CuGaTe$_2$ and CuInTe$_2$, our calculated hole mobilities, electrical conductivities, and Seebeck coefficients agree well with available experimental data~\cite{CuGaTe2,CuInTe2,2020Origin}, indicating that phonon scattering dominates the electrical resistivity. At fixed $T$ and $n_\mathrm{h}$, the hole mobility increases substantially as $X$ varies from S to Te, leading to a corresponding increase in $\sigma$ from S to Se to Te. This trend is driven by progressively weaker polar–optical–phonon scattering, stemming from a reduced ionic contribution to the dielectric response for heavier chalcogens. Combined with smaller transport effective masses, Cu$M$Te$_2$ compounds therefore exhibit high electrical conductivities and large power factors. Our phonon and $\kappa_{\mathrm{L}}$ calculations show that three-phonon processes play a more important role than four-phonon processes in suppressing heat transport in these materials. The weak correlations between $\kappa_{\mathrm{L}}$ and atomic mass, and between $\kappa_{\mathrm{L}}$ and sound velocity, reflect the dominant role of three-phonon scattering. The anomalously low $\kappa_{\mathrm{L}}$ in Cu$M$Se$_2$ also arises from their high three-phonon scattering rates. For the TE performance, our calculated $ZT^{\perp}$ for CuInTe$_2$ (0.71) is in excellent agreement with experiment (0.74)~\cite{2020Origin}. The peak $ZT^{\perp}$ values of CuAlS$_2$ and CuGaS$_2$ are also relatively large, at 0.76 and 0.74, respectively. Among all compounds, the largest $ZT$ occurs in CuAlS$_2$ at 800\,K, primarily due to its low $\kappa_{\mathrm{L}}$ at this temperature. Given the limited band degeneracy and moderate hole mobilities, strategies that further reduce $\kappa_{\mathrm{L}}$ and enhance $\sigma$ are the most effective routes to improving the TE performance of Cu$MX_2$ chalcopyrites.

	\section{ACKNOWLEDGMENTS}
	The authors acknowledge the support of the National Science Foundation of China (Grant No. 12374024) and Fundamental Research Funds for the Central Universities (No. FRF-BRB-25-006). The computing resource was supported by USTB MatCom of Beijing Advanced Innovation Center for Materials Genome Engineering.
	
	\bibliography{ref}

\end{document}